\documentclass{iau}
\usepackage{graphicx}
\def\aj{\textit{AJ}}

\def\apj{\textit{ApJ}}
\def\apjl{\textit{ApJ} (Letters)}

\def\aap{\textit{A\&A}}
\def\araa{\textit{AA\&AR}}
\def\mnras{\textit{MNRAS}}

\def\nat{\textit{Nature}}
\def\nar{\textit{New Astron. Revs}}
\def\ssr{\textit{Space Sci. Revs}}

\title[The impact of AGN] 
{The impact of AGN on their host galaxies}

\author[C.~M. Harrison]   
{C.~M. Harrison$^1$}

\affiliation{$^1$Department of Physics, Durham University, South Road,
  Durham, DH1 3LE, UK \\ email: {\tt c.m.harrison@durham.ac.uk} \\[\affilskip]}

\pubyear{2013}
\volume{304}  
\pagerange{001--007}
\jname{Multiwavelength AGN Surveys and Studies}
\editors{Areg Mickaelian, Felix Aharonian \& David Sanders}
\begin{document}

\maketitle

\begin{abstract}
In these proceedings I briefly: (1) review the impact (or ``feedback'') that active
galactic nuclei (AGN) are predicted to have on their host galaxies and
larger scale environment, (2) review the observational
evidence for or against these predictions and (3) present new results
on ionised outflows in AGN. The observational support for the ``maintenance mode'' of feedback is strong (caveat the details); AGN at the centre of
massive halos appear to be regulating the cooling of hot gas,
which could in turn control the levels of future star formation (SF) and
black hole growth. In contrast, direct observational support for more
rapid forms of feedback, which dramatically impact on SF (i.e., the ``quasar mode''), remains
elusive. From a systematic study of the spectra of $\approx24 000$ AGN we find that extreme
ionised gas kinematics are common, and are most prevalent in radio bright AGN
($L_{{\rm 1.4\,GHz}}>10^{23}$\,W\,Hz$^{-1}$). Follow-up IFU observations
have shown that these extreme gas kinematics are extended over 
kilo-parsec scales. However, the co-existence of high-levels of SF, luminous
AGN activity and radio jets raises interesting questions on the
primary drivers and impact of these outflows. Galaxy-wide, high-mass outflows
are being observed in an increasing number of AGN and are a
plausible mechanism for the depletion of gas; however, there is
still much work to be done to determine the physical processes that drive these
outflows and to measure the level of impact that they have on their host galaxies.
 
\end{abstract}

\firstsection 
\section{Introduction}
Among the most important discoveries in modern astronomy is that all
massive galaxies host a central super-massive black hole (BH). These BHs
primarily grow through mass accretion and become visible as active galactic nuclei (AGN). Surprisingly, observations of galaxies in the local Universe have shown that BH
masses are proportional to that of their host galaxy spheroid (typical mass ratio $\approx1.4\times10^{-3}$; e.g., \cite[Kormendy \& Richstone 1995]{Kormendy95}; \cite[Magorrian
et~al. 1998]{Magorrian98}; \cite[Temaine et~al. 2002]{Tremaine02};
\cite[G\"ultekin et~al. 2009]{Gultekin09}) despite a factor of $\approx$billion difference in their physical size scales. Over the last couple of decades theoretical models of
galaxy formation have implemented energetic ``feedback'' processes
between BH growth and galaxy growth to reproduce this relationship and
many other fundamental observable properties of galaxies, intracluster
medium (ICM) and the intergalactic medium (IGM). Excellent reviews
covering this topic can be found in: \cite[Cattaneo et~al. (2009)]{Cattaneo09};
\cite[Alexander \& Hickox (2012)]{Alexander12}, \cite[McNamara \&
Nulsen (2012)]{McNamara12} and \cite[Fabian (2012)]{Fabian12}. In these proceedings, I briefly summarise some of the key results, discuss
the most recent research and present new results from a systematic study
of ionised outflows in AGN.

\section{Matching models with observations: AGN to the rescue}
\label{Sec:Models}

\noindent AGN are incredible energy sources. During the growth of a BH, huge amounts of energy can be
liberated. Assuming that the energy released during mass
accretion onto a BH is $E_{{\rm BH}}\approx0.1M_{{\rm BH}}c^{2}$, to build a BH with mass $M_{{\rm
    BH}}=10^{8}$\,M$_{\odot}$ would correspond to $E_{{\rm
    BH}}=10^{61}$\,erg. This total accretion energy is two orders of magnitude higher
than the binding energy of the host galaxy spheroid ($M_{{\rm Sph}}\approx10^{11}$\,M$_{\odot}$; $E_{{\rm
    BE}}\approx10^{59}$\,erg) and can be comparable to, or higher than,
the thermal energy of the gas in the dark matter halo (e.g., \cite[Bower et~al. 2008]{Bower08}). If even a small fraction of the
accretion energy can couple to the gas over $\approx$\,0.1--1000\,kpc,
growing BHs have the potential to regulate their own growth and impact
upon the gas in their host galaxies and that in the larger scale
environment. Indeed, to successfully reproduce many of the fundamental
properties of galaxies, ICM and the IGM models of galaxy formation
have found it necessary for AGN to inject some of this energy into the surrounding gas. I discuss some examples of this below.

One of the most famous theoretical results on the impact of AGN comes from
semi-analytical models, which are unable to reproduce the cut-off at
the bright end of the galaxy luminosity function without AGN injecting
energy into the halo gas (e.g., \cite[Bower
et~al. 2006]{Bower06}; \cite[Croton et~al. 2006]{Croton06}). In these
models low accretion rate AGN located in the quasi-hydrostatic halos of massive galaxies (i.e., halo masses
$\gtrsim3\times10^{11}h^{-1}$\,M$_{\odot}$ for local galaxies) efficiently suppress the cooling of hot gas through the
so-called ``maintenance-mode'' or ``radio-mode'', resulting in the
regulation of future star formation (SF) and BH growth in the host
galaxy. This method of feedback has also been used to explain
other observables such as the lack of cold gas in
galaxy clusters (e.g., \cite[Quilis et~al. 2001]{Quilis01}; \cite[Peterson
et~al. 2003]{Peterson03}), the colour bi-modality of galaxies and the evolution of
cosmic SF density (e.g., \cite[Bower et~al. 2006]{Bower06}; \cite[Croton et~al. 2006]{Croton06}). 

The relationship between the X-ray luminosity ($L_{{\rm X}}$) and X-ray temperature ($T_{{\rm X}}$) of
the ICM within groups and clusters is another key observable for
models to reproduce. This is observed to be steeper than expected if
gravity was the only source of heating (e.g., \cite[Markevitch 1998]{Markevitch98}; \cite[Horner 2001]{Horner01}; \cite[Sun et~al. 2009]{Sun09}; \cite[Stott
et~al. 2012]{Stott12}). AGN are thought to be the most viable source of
extra heating (e.g., \cite[Valageas \& Silk 1999]{Valageas99};
\cite[Wu, Fabian \& Nulsen 2000]{Wu00}) and simulations require that
AGN remove some of the low-entropy gas from the centers of halos to
reproduce the observed $L_{{\rm X}}$--$T_{{\rm X}}$ relationship (e.g., \cite[Bower
et~al. 2008]{Bower08}; \cite[Puchwein et~al. 2008]{Puchwein08}; \cite[McCarthy et~al. 2010]{McCarthy10}); however,
the details of this process are yet to be fully understood
(\cite[McCarthy et~al. 2011]{McCarthy11}; \cite[McNamara \& Nulsen
2012]{McNamara12}). 

The ``maintenance-mode'' of feedback, which has dominated the previous
discussion, is thought to be most efficient
in the most massive halos, at late times and is associated with BHs
with low mass accretion rates  (e.g., \cite[Churazov et~al. 2005]{Churazov05}; \cite[Bower et~al. 2006]{Bower06};
\cite[McCarthy et~al. 2011]{McCarthy11}). In contrast, a more rapid/catastrophic form of interaction
between AGN and their host galaxies (sometimes
named the ``quasar mode'') is proposed during high accretion states
(e.g., \cite[Silk \& Rees 1998]{Silk98}). To reproduce many
fundamental observables of massive galaxies and BHs, simulations that invoke this
form of feedback typically require $\approx5-15\%$ of the accretion energy
to couple to the surrounding gas, expel gas through
outflows and consequently suppress or shut down future BH growth and
SF (e.g., \cite[Benson et~al. 2003]{Benson03};
\cite[DiMatteo et~al. 2005]{DiMatteo05}; \cite[Booth \& Schaye 2010]{Booth10}; \cite[DeBuhr
et~al. 2012]{Debuhr12}). Analytical models have also used the idea of galaxy-wide
outflows, initially launched by AGN, to explain the BH-mass--galaxy
mass relationship\footnote{Another interpretation is this is a natural
  consequence of repeated mergers (e.g., \cite[Peng 2007]{Peng07})} (e.g.,
\cite[Fabian 1999]{Fabian99}; \cite[Murray et~al. 2005]{Murray05}; \cite[King et~al. 2011]{King11}; see
review in \cite[Alexander \& Hickox 2012]{Alexander12}). However,
there are doubts about whether AGN-driven outflows are globally a
sufficient form of feedback (e.g., \cite[DeBuhr
et~al. 2010; 2012]{Debuhr10,Debuhr12}) and it is possible that this
``quasar mode'' is necessary to preheat/expel gas at early times before the
``maintenance-mode'' takes over at lower redshifts (\cite[Gabor
et~al. 2011]{Gabor11}; \cite[McCarthy et~al. 2011]{McCarthy11}).

Finally, AGN-driven outflows (in addition to supernovae) may be required to unbind gas from their host galaxies to
fully explain the chemical enrichment of ICM and the IGM
(e.g., \cite[Borgani et~al. 2008]{Borgani08}; \cite[Wiersma
et~al. 2009]{Wiersma09}; \cite[Fabjan et~al. 2010]{Fabjan10}) and it
has also been proposed that, in some cases, AGN-driven outflows could
also cause {\em positive} ``feedback'' by triggering SF 
episodes through induced pressure of the cold gas (e.g.,
\cite[Nayakshin \& Zubovas 2012]{Nayakshin12}; \cite[Ishibasi \&
Fabian 2012]{Ishibashi12}). 

Galaxy formation models have proven to be very successful at
reproducing global observable properties of galaxies, ICM and the IGM. However, they often rely on
diverse, and artificial prescriptions when implementing AGN
``feedback'' , in particular to how the BH accretion energy couples to
the gas. We must therefore appeal to observations to look for {\em direct}
evidence that AGN have an impact on their host galaxies and
larger scale environment and to constrain the details of how, when and where this impact occurs. 

\section{Searching for observational evidence of the impact of AGN}
\label{Sec:Observations}

\begin{figure}
 \vspace*{-0.5 cm}
\begin{center}
 \includegraphics[width=3.2in,angle=90]{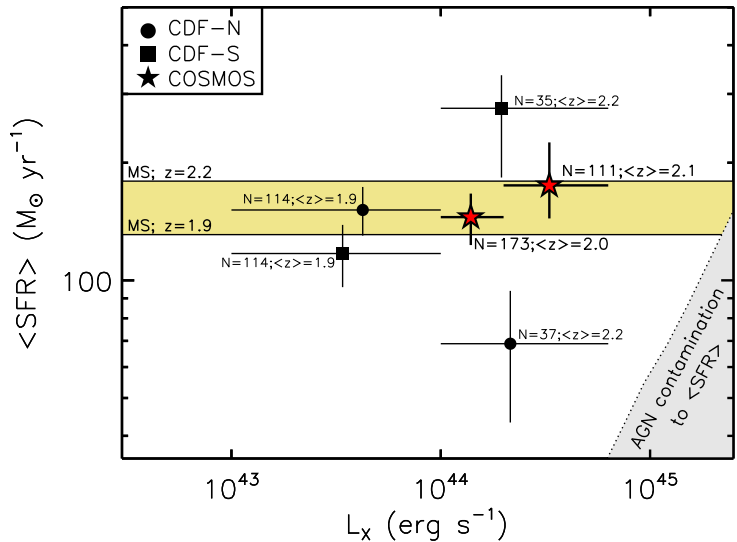} 
\vspace*{-0.5 cm}
 \caption{Mean SFR versus X-ray luminosity for AGN in three fields,
   with the number of sources and mean redshift of each bin
   indicated. The solid horizontal lines show the typical
 SFRs for non-AGN star-forming galaxies of comparable masses
 ($\approx$10$^{10.9}$\,M$_{\odot}$) at
 $z=1.9$ and $z=2.2$ (\protect\cite[Elbaz et al. 2011]{Elbaz11}). The X-ray luminous AGN have the same mean SFRs as
 the non-AGN. Variations at the high-luminosity end, that were previously
 reported in the literature, appear to be due to source statistics and field-to-field
 variations. Adapted from \protect\cite[Harrison et~al. (2012a)]{Harrison12b}.}
   \label{Fig:H12b}
\end{center}
\end{figure}

\noindent As already briefly mentioned, there are several pieces of
indirect observational evidence that AGN are injecting energy into their larger
scale environment inside groups and clusters (e.g., the low rates of
cooling; the steep $L_{{\rm X}}$-$T_{{\rm X}}$ relationship). However,
more direct observational evidence is found through combining X-ray and
radio imaging of clusters where cavities in the X-ray emitting
gas, indicating the prevention of cooling, are observed spatially co-incident with radio sources
(e.g., \cite[Boehringer et~al. 1993]{Boehringer93}; \cite[Carilli
et~al. 1994]{Carilli94}; \cite[McNamara et~al. 2000]{McNamara00}; \cite[Hlavacek-Larrondo
et~al. 2013]{HlavacekLarrondo13}). Most of the injected energy is in
the form of mechanical energy from jets, which inflate these cavities and is
capable of exceeding the X-ray luminosities of their cooling
atmospheres (e.g., \cite[B\^irzan et~al. 2008]{Birzan08};
\cite[Cavagnolo et~al. 2010]{Cavagnolo10}; see \cite[McNamara \& Nulsen 2012]{McNamara12} for a
review). Indeed, the rates of cooling and heating (inferred from X-ray
and radio observations respectively) in radio-luminous AGN appear to be in relatively
close balance in clusters (e.g., \cite[B\^irzan et~al. 2004]{Birzan04};
\cite[Dunn \& Fabian 2008]{Dunn08}) and potentially even in groups and massive
elliptical galaxies out to $z\approx1$ (\cite[Best
et~al. 2006]{Best06}; \cite[Smol{\v c}i{\'c}
et~al. 2009]{Smolcic09}; \cite[Danielson
et~al. 2012]{Danielson12}; \cite[Simpson
et~al. 2013]{Simpson13}). These observations are examples of the
truest form of a ``feedback'' loop (c.f., the ``maintenance mode'';
see Section~\ref{Sec:Models}) where cooling of gas triggers AGN
activity which consequently controls its own fuel supply, in addition to that of SF, by
preventing further cooling (also see G. Tremblay's article, these proceedings).

Most of the above focuses on radio AGN in dense environments which are
radiatively inefficient (i.e., low-accretion states) and
may represent a distinct class of AGN compared to their radiatively efficient counterparts
(\cite[Hickox et~al. 2009]{Hickox09}; \cite[Best \& Heckman
2012]{Best12}). Next I consider observations, particularly at high
redshift, that investigate the impact of the radiatively efficient (here-after ``luminous'') AGN population,
which are proposed to be responsible for ``quasar mode'' feedback (Section~\ref{Sec:Models}). Here we are looking
for observational signatures that luminous AGN expel gas from their
host galaxies, redistribute metals and/or impact on SF (Section~\ref{Sec:Models}). Energetic and high-mass
outflows do exist in luminous AGN (Section~\ref{Sec:Outflows});
however, observationally it is not yet clear what {\em impact} that these AGN have on the
formation and evolution of galaxies. A lot of recent observational work has concentrated on investigating the star formation rates (SFRs) in the host
galaxies of AGN compared to the overall population (see \cite[Alexander \& Hickox et~al. 2012]{Alexander12};
\cite[Fabian 2012]{Fabian12} for other examples). Subsequent to the surveys carried out by the {\em Herschel}
satellite, the study of SFRs of AGN has
increased greatly, largely because the far-infrared wavelengths observed ($\lambda=70-500\mu m$) arguably provide the most reliable SFR
measurements in luminous AGN (e.g., \cite[Mullaney
et~al. 2012]{Mullaney12}). {\em Herschel} results show that, on average, the SFRs of X-ray selected AGN with moderate luminosities (i.e., $L_{\rm
  X}=10^{42}-10^{44}$\,erg) are consistent with non-AGN star-forming galaxies of the same mass
and redshift (e.g., \cite[Shao et~al. 2010]{Shao10}; \cite[Mullaney
et~al. 2012]{Mullaney12a}) in agreement with other work (e.g., \cite[Xue
et~al. 2010]{Xue10}; \cite[Lutz et al. 2010]{Lutz10}; \cite[Mainieri
et~al. 2011]{Mainieri11}). However, the mean SFRs of the most luminous X-ray AGN ($L_{\rm
  X}>10^{44}$\,erg\,s$^{-1}$) is a more controversial topic pre- and post-{\em
  Herschel}, with claims of both enhanced and suppressed SFRs
compared to the overall population (e.g., \cite[Lutz et~al. 2010]{Lutz10};
\cite[Rovilos et~al. 2012]{Rovilos12};
\cite[Page et~al. 2012]{Page12}). However, a major factor behind these different conclusions
is poor source statistics and/or
field-to-field variations (\cite[Harrison
et~al. 2012a]{Harrison12b}; see Fig.~\ref{Fig:H12b}). When looking at
large samples of X-ray AGN the mean SFRs are consistent with the overall star-forming population out to
$L_{{\rm x}}\approx10^{45}$\,erg\,s$^{-1}$, at least over the redshift
range $z=$\,1--3 (see Fig.~\ref{Fig:H12b}; see \cite[Rosario
et~al. 2012]{Rosario12}). Taken at face-value, these results might
appear to suggest that luminous AGN have no impact on SF; however, there are many factors to
consider. For example, depending on the relative timescales and variability of X-ray
luminous AGN activity compared to SF episodes (see \cite[Hickox et~al. 2013]{Hickox13})
any subtle signatures of SFR suppression or enhancement in AGN could be very challenging
to detect using these methods (\cite[Harrison et~al. 2012a]{Harrison12b}).

Finally, directly conflicting studies of different classes of AGN
(e.g., radio luminous, X-ray luminous) have recently arisen in the
literature, arguing that each class of AGN are capable of suppressing, enhancing, or having no impact
on SF (e.g., \cite[Kalfountzou et~al. 2012]{Kalfountzou12}; \cite[Zinn et~al. 2013]{Zinn13}; \cite[Karouzos et~al. 2013]{Karouzos13};
\cite[Feltre et~al. 2013]{Feltre13}; \cite[Rosario et~al. 2013]{Rosario13b}). There are potentially many explanations for
this. Sample selection is likely to be a key factor, for example,
constructing samples based on shallow radio or infrared
surveys will only identify objects with very high SFRs. Also, the various
approaches for selecting AGN at different wavelengths (i.e.,  X-rays, optical,
mid-infrared or radio), all have different issues of contamination
(from non-AGN) and completeness and may even result in selecting AGN of different masses, evolutionary stages or environments (\cite[Hickox
et~al. 2009]{Hickox09}; \cite[Goulding et~al. 2013]{Goulding13}). We
must first understand the many observational biases and obtain SFR
distributions, down to reasonable limits, for complete samples before we can draw firm conclusions on the impact of luminous AGN on SF.

\section{Galaxy-wide outflows: a ``lever-arm'' for AGN to impact on
  their host galaxies}
\label{Sec:Outflows}
\begin{figure}
 \vspace*{-0.1 cm}
\begin{center}
 \includegraphics[width=2.4in,angle=0]{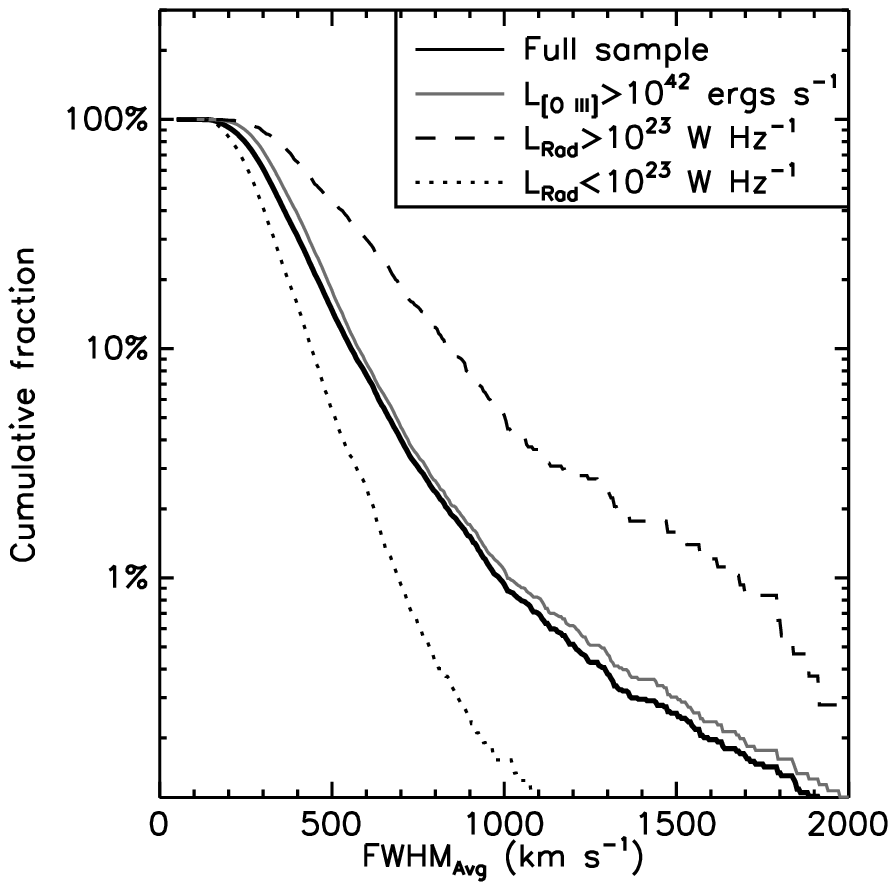} 
\includegraphics[width=2.6in,angle=0]{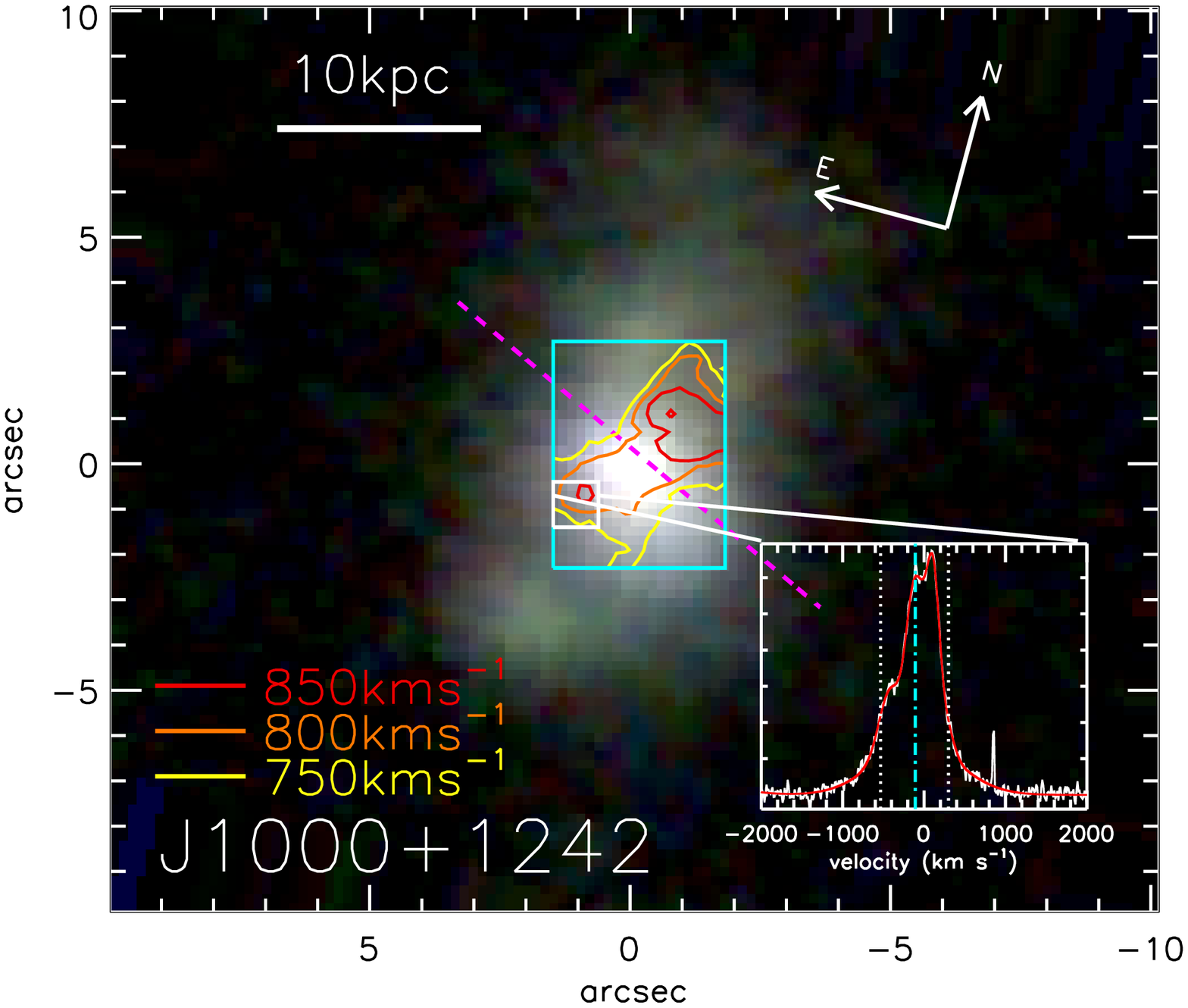} 
 \vspace*{-0.2 cm}
 \caption{{\em Left:} The fraction of $z<0.4$ AGN with [O~{\sc iii}]
   FWHM$_{{\rm Avg}}$ (weighted average of broad and narrow components)
   greater than the given values. Values of FWHM$_{{\rm Avg}}>1000$\,km\,s$^{−1}$ are $\approx$5 times more
   prevalent among AGN with L$_{{\rm 1.4 GHz}}>10^{23}$\,W\,Hz$^{−1}$ (\protect\cite[Mullaney
   et~al. 2013]{Mullaney13}). {\em Right:} The SDSS image of an
   example AGN from our follow-up study of the \protect\cite[Mullaney et~al. (2013)]{Mullaney13}
   parent sample (Harrison et~al. in prep). Our IFU data (contours)
   reveal a high-velocity bi-polar ionised outflow, perpendicular to the galactic
   kinematic major axis (dashed line). The inset shows an [O~{\sc iii}]
   emission-line profile from the indicated region. Kiloparsec-scale
   extreme ionised gas kinematics are observed in all of the
   sixteen AGN with IFU data (Harrison et~al. in prep).}
   \label{Fig:outflows}
\end{center}
\end{figure}

\noindent AGN-driven, galaxy-wide outflows are a key prediction of
many models (Section~\ref{Sec:Models}). AGN outflows are known to be
common on $\lesssim1$~pc scales since very high-velocity winds
(up to $\approx$~0.1\,c) are observed in a large
fraction of AGN and may even be ubiquitous (e.g.,\ \cite[Pounds et~al. 2003]{Pounds03};
\cite[Ganguly \& Brotherton 2008]{Ganguly08}
\cite[Tombesi et~al. 2010]{Tombesi10}; see \cite[Fabian 2012]{Fabian
  et~al. 2012} for review). However, while significant, these studies
provide little direct insight into what effect these outflows have on
the gas and SF over galactic scales. This requires spatially-resolved
kinematic measurements of ionised gas, molecular gas and
atomic gas. Here I concentrate on ionised outflows which currently provide the easiest means to perform statistical
studies (for observations in other phases see S. Veilleux, these
proceedings). Broad ($>$500--1000\,km\,$^{-1}$), high-velocity and
spatially extended [O~{\sc iii}]5007 emission is one key diagnostic
that has revealed galaxy-wide ionised outflows in integral field spectroscopy (IFU data) of low and high redshift
AGN (e.g., \cite[Nesvadba et~al. 2008]{Nesvadba08}; \cite[Ruke
\& Veilleux 2011]{Rupke11}; \cite[Harrison et~al. 2012b]{Harrison12a};
\cite[Liu et~al. 2013]{Liu13}). These IFU observations have
demonstrated that galaxy-wide ionised outflows exist and have the
potential to drive gas out of their host galaxies; however, they are
typically of small and inhomogeneous samples of
AGN. Key questions that arises from these studies are: ``how representative are the objects of
the overall population?'' and ``where are ionised outflows most
preferentially found?''. 

To address the questions above we performed emission-line 
de-composition (fitting broad and narrow components) on $\approx$24 000 $z<0.4$ AGN from the SDSS
survey (\cite[Mullaney et~al. 2013]{Mullaney13}). We consequently measured the prevalence of ionised
outflow features in the overall AGN population and developed a well constrained parent sample for the basis for detailed follow-up
observations. After accounting for the known
correlation between bolometric luminosity ($L_{{\rm  AGN}}$) and radio
luminosity ($L_{{\rm 1.4\,GHz}}$), we found that the most extreme line widths
are preferentially found in radio bright systems
($L_{{\rm 1.4\,GHz}}>10^{23}$\,W\,Hz$^{-1}$;
see Fig.~\ref{Fig:outflows}) whilst no clear trends are observed with $L_{{\rm  AGN}}$ or
Eddington ratio. Follow-up IFU
observations of sixteen of the luminous AGN ($L_{{\rm
    [O~III]}}>10^{41.7}$\,erg\,s$^{-1}$; Harrison et~al. in prep) have shown that the
high-velocity features, identified in \cite[Mullaney
et~al. (2013)]{Mullaney13}, are extended over $\gtrsim$(6--16)\,kpc in
all cases, and show a range of morphology and structure (e.g.,
see Fig.~\ref{Fig:outflows}). We measure outflow properties, SFRs and AGN luminosities and search for evidence of radio jets in the
sample. The implied mass outflow rates are comparable to the host
galaxy SFRs [i.e., $\lesssim$\,(7--100)\,M$_{\odot}$\,yr$^{-1}$] and
the energy injection rates are high
($\approx10^{41-43}$\,erg\,s$^{-1}$) in broad agreement with predictions of the ``quasar
mode'' of feedback (Section~\ref{Sec:Models}). The observed AGN have
radio luminosities of $L_{{\rm
    1.4\,GHz}}\gtrsim10^{23}$\,W\,Hz$^{-1}$; however, their radio emission is from a mixture of SF and AGN
activity, raising the interesting question: which physical process is
responsible for the most extreme ionised gas kinematics? Full results
and discussion of the IFU sample will soon be presented in Harrison
et~al. (in prep).

As a final note, it is worth remembering that the {\em
  existence} of galaxy-wide outflows is not direct proof that they have a
long-term impact on their host galaxies. Even the most massive outflows travelling at the
galaxy escape velocity may stall in the galaxy halo, re-collapse and cool at later times (along with new fuel supplies),
resulting in the re-ignition of SF and BH growth (e.g., \cite[Harrison
et~al. 2012b]{Harrison12b}; \cite[Lagos et~al. 2008]{Lagos08}; \cite[McCarthy
et~al. 2011]{McCarthy11}; \cite[Gabor et~al. 2011]{Gabor11};
\cite[Rosas-Guevara et~al. 2013]{RosasGuevara13}). We have made great progress in identifying and
characterising outflows in AGN; however, there is crucial
work still to be done to measure the impact that galaxy-wide outflows
have on their host galaxies.

\end{document}